\documentclass[prd,reprint,nofootinbib,showpacs,superscriptaddress]{revtex4-1}
\usepackage{graphicx} 
\usepackage{hyperref}
\usepackage{amsfonts}
\usepackage{amsmath,amssymb}
\usepackage{bm} 
\usepackage{color}
\usepackage{epstopdf} 
\usepackage{epsfig}
\usepackage{subfig} 
\usepackage{float}

{\rm }

\def\be{\begin{equation}}
 \def\ee{\end{equation}}
 \def\bea{\begin{eqnarray}}
 \def\eea{\end{eqnarray}}
 \def\bes{\begin{eqnarray}}
 \def\ees{\end{eqnarray}}
 \def\bi{\begin{itemize}}
 \def\ei{\end{itemize}} 

\def\2{\frac{1}{2}}
\def\4{\frac{1}{4}}

\begin{document}

\title{Noise analysis of simultaneous quantum key distribution and classical communication scheme using a true local oscillator}

\author{Bing Qi}
\email{qib1@ornl.gov}
\affiliation{Quantum Information Science Group, Computational Sciences and Engineering Division,
Oak Ridge National Laboratory, Oak Ridge, TN 37831, USA}
\affiliation{Department of Physics and Astronomy, The
University of Tennessee, Knoxville, TN 37996, USA
}

\author{Charles Ci Wen Lim}
\email{elelimc@nus.edu.sg}
\affiliation{Quantum Information Science Group, Computational Sciences and Engineering Division,
Oak Ridge National Laboratory, Oak Ridge, TN 37831, USA}
\affiliation{Department of Electrical and Computer Engineering, National University of Singapore, 117583, Singapore}

\date{\today}

\begin{abstract}

Recently, we proposed a simultaneous quantum and classical communication (SQCC) protocol, where random numbers for quantum key distribution (QKD) and bits for classical communication are encoded on the \emph{same} weak coherent pulse, and decoded by the \emph{same} coherent receiver. Such a scheme could be appealing in practice since a single coherent communication system can be used for multiple purposes. However, previous studies show that the SQCC protocol can only tolerate very small phase noise. This makes it incompatible with the coherent communication scheme using a true local oscillator (LO), which presents a relatively high phase noise due to the fact that the signal and the LO are generated from two independent lasers. In this paper, we improve the phase noise tolerance of the SQCC scheme using a true LO by adopting a refined noise model where phase noises originated from different sources are treated differently: on one hand, phase noise associated with the coherent receiver may be regarded as \emph{trusted} noise, since the detector can be calibrated locally and the photon statistics of the detected signals can be determined from the measurement results; on the other hand, phase noise due to the instability of fiber interferometers may be regarded as \emph{untrusted} noise, since its randomness (from the adversary's point to view) is hard to justify. Simulation results show the tolerable phase noise in this refined noise model is significantly higher than that in the previous study where all the phase noises are assumed to be untrusted. We conduct an experiment to show the required phase stability can be achieved in a coherent communication system using a true LO. \footnote{This manuscript has been authored by UT-Battelle, LLC under Contract No. DE-AC05-00OR22725 with the U.S. Department of Energy. The United States Government retains and the publisher, by accepting the article for publication, acknowledges that the United States Government retains a non-exclusive, paid-up, irrevocable, world-wide license to publish or reproduce the published form of this manuscript, or allow others to do so, for United States Government purposes. The Department of Energy will provide public access to these results of federally sponsored research in accordance with the DOE Public Access Plan (http://energy.gov/downloads/doe-public-access-plan).
}

\end{abstract}

\maketitle

\section{Introduction}
\label{sec:1}

Quantum key distribution (QKD) allows two remote parties, traditionally called Alice and Bob, to generate a secure key through an insecure quantum channel fully controlled by an adversary (Eve) \cite{BB84,E91,Gisin02,Scarani09,Lo14,Diamanti16}. The secure key can be further applied in other cryptographic protocols to enhance communication security.

One of the major roadblocks in the wide adoption of QKD is the high cost: dedicated communication infrastructures (such as dark fibers) and expensive devices (such as single photon detectors), are commonly required in today's commercial QKD systems. It is thus imperative to come up with cost-effective QKD solutions. Recently, in light of the similarity between continuous-variable (CV) QKD based on coherent detection \cite{GMCS} and classical coherent communication, we proposed a simultaneous quantum and classical communication (SQCC) protocol where Gaussian distributed random numbers for QKD and bits for classical communication are encoded on the \emph{same} weak coherent pulse, and decoded by the \emph{same} coherent receiver \cite{Qi16}. Since a single coherent communication system can be used for both classical communication and QKD, it can effectively reduce the cost of QKD itself.

However, previous studies show the SQCC protocol can only tolerate very small phase noise \cite{Qi16}. This is mainly due to the cross-talk between the QKD signal and classical communication signal: on one hand, the random QKD signal appears as an additional noise source in the classical communication. To ensure the classical bit error rate (BER) is below a given threshold, a larger modulation amplitude of the classical signal would be required comparing with the case of conducting classical communication alone. On the other hand, since the QKD signal is superimposed on the classical signal, the variance of excess noise due to phase fluctuation is proportional to the power of the classical signal. A larger modulation amplitude of the classical signal will result in a higher excess noise in QKD, hence resulting in a poorer performance. To achieve high-performance classical communication and QKD at the same time, the tolerable phase noise variance is less than $10^{-4} rad^2$ in the previous study \cite{Qi16}. Experimentally, phase noise below $10^{-4} rad^2$ has been demonstrated in CV-QKD experiments using a distributed local oscillator (LO), where the LO for coherent detection is generated from Alice's signal laser and distributed to Bob through an insecure quantum channel \cite{Lodewyck07, Qi07, Jouguet13, Huang16, ZLC17}. However, it could be difficult to achieve such a small phase noise in CV-QKD using a \emph{true} LO, where the LO is generated by Bob using an independent laser source. Note that CV-QKD using a true LO is very appealing in practice due to its simple design and enhanced security \cite{Qi15,Soh15,Huang15,Kleis17,Corvaja17}.

Can we relax the requirement of very low phase noise in the SQCC protocol? In QKD, Alice and Bob can quantify the information gained by Eve from the observed noise and other system parameters: a higher noise level implies more information gained by Eve thus a lower secure key rate. One conservative approach to deal with noise in QKD is to assume all the observed noises are due to Eve's attack. This approach may overestimate Eve's information since practical QKD systems present \emph{intrinsic} noises not necessarily controllable by Eve. An alternative approach is to assume that certain intrinsic noises well protected from Eve are \emph{trusted} in the security proof. This approach can typically lead to a better QKD performance. For example, the trusted detector noise model has been widely adopted in long-distance CV-QKD experiments \cite{GMCS, Lodewyck07, Qi07, Jouguet13, Huang16, Kumar15}. More recently, the trusted source noise model has also been studied in CV-QKD \cite{Usenko10, Shen11, Jouguet12, Yang12, Usenko16}.

In this paper, we improve the phase noise tolerance of the SQCC scheme using a true LO by adopting a refined noise model where phase noises originated from different sources are treated differently: on one hand, phase noise associated with the coherent receiver may be regarded as \emph{trusted} noise, since the detector can be calibrated locally and the photon statistics of the detected signals can be determined from the measurement results. This is consistent with the commonly adopted assumption of trusted detector noise in practical CV-QKD; on the other hand, phase noise due to the instability of fiber interferometers are regarded as \emph{untrusted} noise, since its randomness (from Eve's point to view) is hard to justify. We conduct numerical simulations of the SQCC protocol using a true LO based on the above noise model. Simulation results show the tolerable phase noise in this refined noise model is significantly higher than that in the previous study where all the phase noise are assumed to be untrusted. Based on a design proposed in \cite{Marie17}, we conduct an experiment to show the required phase stability can be achieved in a coherent communication system using a true LO generated at Bob's end. Our findings suggest that the SQCC protocol could be a viable solution in practice.

This paper is organized as follows: In Section \ref{sec:2}, we present details of the SQCC protocol based on conjugate homodyne detection. In Section \ref{sec:3}, we develop the noise model of the SQCC using a true LO, and present simulation results based on practical system parameters. In Section \ref{sec:4}, we conduct an experiment to show the required phase stability can be achieved in a coherent communication system using a true LO. Finally, we conclude this paper with a discussion in Section \ref{sec:5}.
 
\section{Protocols}
\label{sec:2}

The QKD protocol adpoted in this paper is the Gaussian-modulated coherent states (GMCS) protocol \cite{GMCS} based on conjugate homodyne detection \cite{Weedbrook04}. We further assume that the QKD protocol is implemented with a true LO generated by Bob, as proposed in \cite{Qi15}. Since conjugate homodyne detection allows Bob to measure both the X-quadrature and the P-quadrature simultaneously, we adopt the quadrature phase shift keying (QPSK) modulation for classical communication. This is quantitatively different from \cite{Qi16}, where binary phase shift keying modulation is used for classical communication.

\subsection{Classical QPSK scheme}

In QPSK, Alice encodes two classical bits, $m_A$ and $n_A$, into the X-quadrature and the P-quadrature of a coherent state, given by
\bes\label{eq1} \vert\psi\rangle=\vert(e^{-im_A\pi}+ie^{-in_A\pi})\alpha)\rangle \ees 
where $\alpha$ is assumed to be a real number. The average photon number $\mu$ of the coherent state $\vert\psi\rangle$ is given by $\mu=2\alpha^2$.

Bob measures both the X-quadrature and the P-quadrature of the incoming signal and uses the signs of the measurement results to decode $m_A$ and $n_A$, i.e., if the measured quadrature value is positive (negative), the corresponding classical bit is assigned as ``0'' (``1''). 

\begin{figure}[t]
	\includegraphics[width=.45\textwidth]{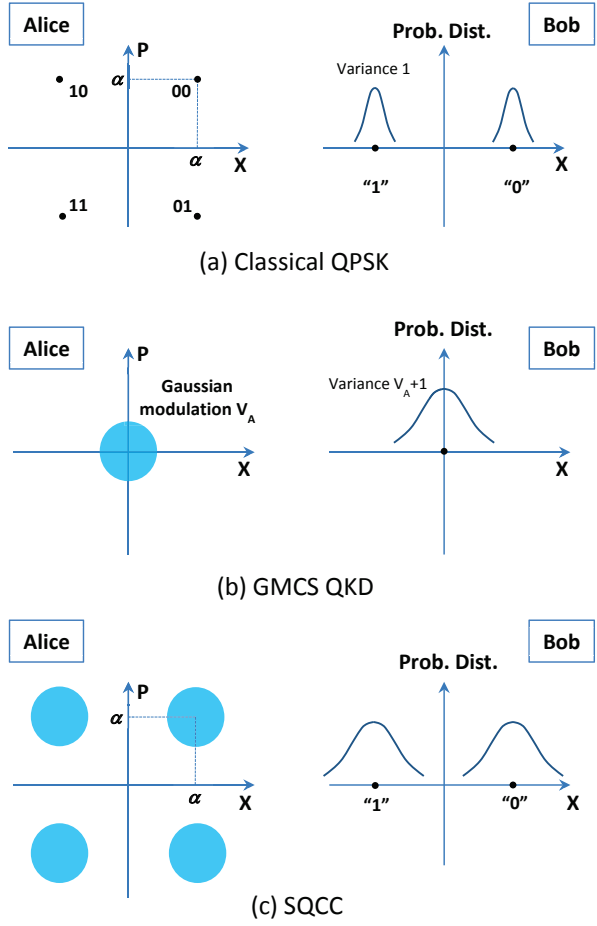}
	\captionsetup{justification=raggedright,
					singlelinecheck=false }
	\caption{Phase space representations of various coherent communication schemes. (a) Classical QPSK scheme. (b) The GMCS QKD scheme. (c) The SQCC protocol. The figures on the right show the probability distributions of X-quadrature measurement.} 
	\label{fig:1}
\end{figure}

\subsection{The GMCS QKD based on conjugate homodyne detection}

In GMCS QKD based on conjugate homodyne detection \cite{Weedbrook04}, Alice prepares a coherent state $|x_A+ip_A\rangle$, where $x_A$ and $p_A$ are Gaussian random numbers with zero mean and a variance of $V_A N_0$. Here $N_0$ = 1/4 denotes the shot-noise variance. In this paper, all the noise variances are defined in the shot-noise unit. At Bob's end, he performs conjugate homodyne detection to measure both X and P quadratures simultaneously. After repeating the above quantum state transmission and detection process many times, Alice and Bob perform data post-processing. Through an authenticated classical channel, Alice and Bob compare a subset of their data to estimate the transmission efficiency and the noise variance for each quadrature. If the observed noise is below a certain threshold, Alice and Bob can further work out a secure key by performing reconciliation and privacy amplification. See more details in Section \ref{sec:3}.

\subsection{The SQCC protocol}
\label{sec2:C}
In the SQCC protocol, Alice encodes her classical bits $\lbrace m_A, n_A\rbrace$ and Gaussian random numbers $\lbrace x_A, p_A\rbrace$ on a coherent state $|(x_A+e^{-im_A\pi}\alpha)+i(p_A+e^{-in_A\pi}\alpha)\rangle$ and transmits it to Bob, who performs conjugate homodyne detection to measure both X and P quadratures simultaneously.

Bob determines the classical bits $\lbrace m_B, n_B\rbrace$ from the signs of his measurement results $\lbrace x_R, p_R\rbrace$: if $x_R (p_R)>0$, then the bit value of $m_B (n_B)$ is assigned as ``0''. Otherwise, the bit value is assigned as ``1''. To decode Alice's random numbers for QKD, Bob processes his measurement results using on the overall transmittance $T \eta$ and the classical bits $\lbrace m_B, n_B\rbrace$ determined above:
\bes\label{eq2} x_B &=&  \sqrt{\dfrac{2}{T\eta}}x_R+(2m_B-1)\alpha \nonumber\\
p_B &=& \sqrt{\dfrac{2}{T\eta}}p_R+(2n_B-1)\alpha \ees
where $T$ is the channel transmittance, $\eta$ is the detector efficiency, and the factor $\sqrt{2}$ is due to conjugate homodyne detection.

Alice and Bob can further perform data post-processing and work out a secure key from raw keys $\lbrace x_A, x_B\rbrace$ and $\lbrace p_A, p_B\rbrace$, just as in the case of conventional GMCS QKD \cite{GMCS}. The phase-space representations of the above three protocols are shown in Fig.~1.

\subsection{CV-QKD using a true LO}

In all protocols discussed above, an LO is needed in coherent detection. In most existing implementations of CV-QKD, to reduce the phase noise, both the signal and the LO are generated by Alice from the same laser and sent through the insecure quantum channel \cite{GMCS, Lodewyck07, Qi07, Jouguet13, Kumar15, Huang16, ZLC17}. This arrangement, however, may allow Eve to launch sophisticated attacks by manipulating the LO \cite{HML08, Ma13, Huang13, Jouguet133, Huang14}. It also requires complicated multiplexing and demultiplexing schemes to effectively separate the strong LO from the weak quantum signal at the receiver's end. To solve the above problems, CV-QKD using a true LO generated at Bob's side has been developed \cite{Qi15,Soh15}. The scheme presented in \cite{Qi15} works as follows: for each transmission, Alice sends out both a quantum signal and a phase reference pulse generated from the same laser. The quantum signal carries Alice's random numbers, while the phase reference pulse is not modulated. At Bob's end, he performs conjugate homodyne detection on both the quantum signal and the phase reference pulse using two separate LOs generated from his own LO laser. The measurement results from the phase reference pulse are used to recover the phase relation $\phi$ between the two lasers. Using this phase information Bob can classically correct his measurement results of the quantum signal in the post-processing stage by performing the following rotation
\bes\label{eq3} x'_R &=& x_R cos\phi-p_R sin\phi \nonumber\\
p'_R &=& x_R sin\phi+p_R cos\phi. \ees

Various schemes have been proposed to implement CV QKD using a true LO. In Ref.~\cite{Qi15}, the QKD signal and the phase reference pulse are generated by using an amplitude modulator to modulate the outptut of a continuous-wave (cw) laser twice, as shown in Fig.~2(a). Two LO pulses are generated from Bob's laser in the same way. Given Bob's detector noise is much smaller than the shot noise, the main phase noise of this scheme can be estimated by \cite{Qi15}
\bes\label{eq4}\sigma=\dfrac{\Delta t}{\tau_1}+\dfrac{\Delta t}{\tau_2}+\dfrac{2N_0}{\eta n_{ref}} \ees 
where $\Delta t$ is the time delay between the signal pulse and the phase reference pulse, $\tau_1$ ($\tau_2$) is the coherent time of the signal (LO) laser, and $n_{ref}$ is the average photon number of the phase reference pulse at Bob's side.

We define $\sigma_B=\dfrac{2N_0}{\eta n_{ref}}$. It represents the shot-noise contribution, and in principle can be suppressed by using a strong phase reference pulse. The first two terms on the RHS of (4) are fundamental phase noises associated with the finite linewidth of the lasers, which could be reduced by decreasing the time delay $\Delta t$ or using lasers with longer coherent time (narrower linewidth).

In \cite{Marie17}, Marie and All\'{e}aume proposed a modified scheme where the signal pulse and the phase reference pulse are split from a common pulse using a path-unbalanced interferometer, as shown in Fig.~2(b). Since this scheme can effectively remove the phase noise contributed by the lasers, the residual phase noise is mainly determined by the last term on the RHS of (4) and the phase instability of the path-unbalanced interferometers. We adopt this modified scheme in this paper.  

\begin{figure}[t]
	\includegraphics[width=.5\textwidth]{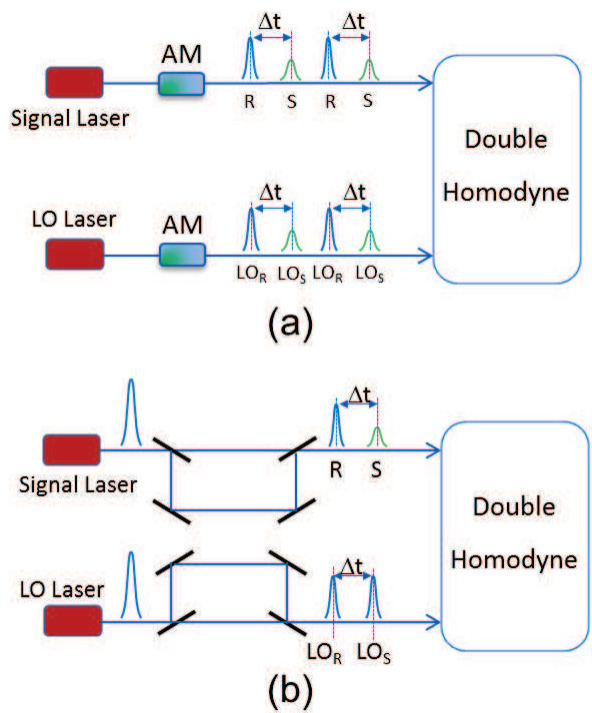}
	\captionsetup{justification=raggedright,
					singlelinecheck=false }
	\caption{Two different ways to generate phase reference pulses in CV-QKD using a true LO. (a) The QKD signal (S) and the phase reference pulse (R) are generated from a continuous-wave laser by using an amplitude modulator (AM) \cite{Qi15}. (b) The QKD signal and the phase reference pulse are split from a common laser pulse using a path-unbalanced interferometer \cite{Marie17}.}
	\label{fig:2}
\end{figure}

\section{Noise analysis}
\label{sec:3}

The performance of the SQCC protocol depends on the noises presented in the system. In this Section, we first present the noise model adopted in this paper, followed by calculations of the BER in classical communication, the secure key rate in QKD, and simulation results based on realistic parameters.

\subsection{Noise model}

The main noise sources considered here are: 1) phase noise in a coherent communication system using a true LO; 2) noise $\varepsilon_{le}$ due to the leakage from the phase reference pulse to the signal; 3) detector noise denoted by $\upsilon_{el}$; 4) signal-independent noise $\varepsilon_0$ from channel and other unidentified or unprotected sources; and 5) vacuum noise. All the noises are assumed to be Gaussian, and we use the same symbol to represent both the noise and its variance. 

In this paper, we adopt the trusted detector noise model by assuming both the detector efficiency $\eta$ and detector noise $\upsilon_{el}$ are well calibrated and out of Eve's control. In contrast, items 2) and 4) are untrusted noises and contributed to Eve's attack. Item 1) is more complicated and can be further separated into two terms: the phase noise $\sigma_I$ due to the instability of the path-unbalanced interferometers (see Fig.~2(b)), and the phase noise $\sigma_B$ given by the last term on the RHS of (4). As discussed in \cite{Marie17}, previous experimental demonstrations of CV-QKD using a distributed LO have shown that phase noise $\sigma_I$ associated with the path-unbalanced interferometers can be very small. For example, phase noises in the order of $10^{-4}\sim10^{-5} rad^2$ have been demonstrated in \cite{Lodewyck07, Huang16}. In comparison, phase noise $\sigma_B$ is typically much higher ($10^{-3} rad^2$ to be shown in this paper).

One crucial assumption we make in this paper is that the phase noise $\sigma_B$ is trusted. Since $\sigma_B$ is determined by the detector and the photon number of the phase reference pulse received by Bob, to justify this assumption in practice, Bob may need to calibrate the detector and the phase reference pulse in the QKD process \footnote{In \cite{Ren17}, the authors discuss an interesting attack in CV QKD using a true LO where Eve intentionally increases the photon numebr of the phase reference pulse receievd by Bob. She could achieve this by conveying the phase refence pulse from Alice to Bob using a lossless channel. If Alice and Bob blindly apply the trusted phase noise model without monitoring the photon number of the phase reference pulse at Bob, they may overestimate the amount of trusted phase noise, and leave room for Eve to hide her attack on the quantum signal. Of course, this attack won't work if Alice and Bob assume all the phase noises are untrusted, as in \cite{Qi15,Soh15}.}. The detector calibration is also required in the trusted detector noise model and has been studied previously \cite{Jouguet133}. Here we present a brief discussion on the calibration of the photon number of the phase reference pulse. As discussed in Section II D, to determine the phase relation between the LO laser and the signal laser, Bob performs conjugate homodyne detetcion to measure both the X-quadrature and P-quadrature of the phase reference pulse. From his measurent results, Bob can also determine the photon statistice of phase reference pulses since classically the quantity $z=X^2+Y^2$ is propotional to the intensity of the phase refrence pulse \cite{QL17}. So, the same measurement device for the LO phase recovery can also be used for phase reference pulse calibration. Given the finite photon number of the phase reference pulse, the corresponding phase noise is essentially originated from vacuum noise, which is truly random to both the QKD users and Eve. We thus assume $\sigma_B$ as trusted noise.

Can we assume the phase noise $\sigma_I$ is also trusted? At first sight, since Eve cannot access the QKD system, it seems reasonable  to assume that $\sigma_I$ cannot be manipulated by Eve. However, the unpredictability of this noise (from Eve's point of view) is hard to justify. If there are some internal patterns of the inteferometer phase drift which are ignored by QKD users but known by Eve, she may compensate this phase drift when the signal propagates through the channel and thus reduce the phase noise. In the mean time, she can attack the quantum signal to gain information at the cost of introducing noise. If the total noise (including the reduced phased noise and the noise due to Eve's attack) equals to the phase noise expected by the users (when Eve does not compensate the phase drift in the channel), Eve's attack cannot be detected. For this reason, we assume $\sigma_I$ is untrusted.

The term $\varepsilon_{le}$ quantifies the noise due to the leakage from the phase reference pulse to the signal. Since the effect of leakage is implementation-specific, we conduct a detailed analysis in Appendix A based on the design to be presented in Section IV, where both time multiplexing and polarization multiplexing are employed to reduce the leakage. As shown in Appendix A, the excess noise contributed by the leakage referred to the input of the channel is given by
\bes\label{eq5} \varepsilon_{le}=\dfrac{n_{ref}\Delta t}{T N_0 \tau_c} \times 10^{-\frac{\xi_A}{10}}\times 10^{-\frac{\xi_P}{10}},\ees
where $\Delta t$ is the time delay between the signal pulse and the phase reference pulse, $\tau_c$ is the coherent time of Alice's laser, and $\xi_A$ and $\xi_P$ are the extinction ratios (in dB) of the amplitude modulator and the polarization multiplexing scheme.

From (5), the excess noise $\varepsilon_{le}$ can be effectively suppressed by improving the extinction ratio $\xi_A$ or $\xi_P$. While amplitude modulators with 65dB extinction ratio have been demonstrated experimentally \cite{Liu17} and applied in CV-QKD experiment \cite{HLW15}, most standard commercial products can achieve an extinction ratio in the range of 20 to 50dB. In the simulation below, we assume an extinction ratio of 30dB for both amplitude modulation and polarization multiplexing. 

\subsection{Bit error rate in classical communication}

In the SQCC protocol, the Gaussian modulation for QKD appears as a Gaussian noise for classical communication. Furthermore, the contribution of phase noises is proportional to the power of classical signal and can be described by $\dfrac{\alpha^2}{N_0}(\sigma_I+\sigma_B)$. The overall noise variance at the receiver's end is given by
\begin{equation}
\begin{split}
N_{tot}=\frac{1}{2}T\eta[V_A+\varepsilon_{le}+\varepsilon_0+\dfrac{\alpha^2}{N_0}(\sigma_I+\sigma_B)] \\
+1+\upsilon_{el},
\end{split}
\end{equation}
where the factor $\frac{1}{2}$ is due to conjugate homodyne detection, since the received signal is split by Bob into two using a symmetric beam splitter.

We assume the channel between Alice and Bob is telecom fiber with an attenuation coefficient of $\gamma$, which is assumed to be 0.2 dB/km. The channel transmittance is given by
\bes\label{eq7} T=10^{\frac{-\gamma L}{10}}\ees
where $L$ is the fiber length in kilometers.

Given the signals transmitted by Alice are described by Eq.~(\ref{eq1}), the BER of the classical QPSK is given by
\bes\label{eq8} C_{BER}=\dfrac{1}{2}{\rm{erfc}}(\dfrac{\sqrt{T\eta}\alpha}{\sqrt{4N_{tot}N_0}})\ees
where erfc($\cdot$) denotes the complementary error function. 

To achieve a BER of $C_{BER}$ in the classical communication, the required displacement $\alpha$ can be determined from Eqs.~(6)--(8) as
\bes\label{eq9} \alpha=w\dfrac{\sqrt{T\eta(V_A+\varepsilon_{le}+\varepsilon_0)+2+2\upsilon_{el}}}{\sqrt{T\eta(2-4w^2\sigma_I -4w^2\sigma_B)}}\ees
where $w$ is defined as
\bes\label{eq10} w={\rm{erf}}^{-1}(1-2C_{BER}). \ees
Here ${\rm{erf}}^{-1}(\cdot)$ is the inverse error function.

We remark that to achieve a BER of $10^{-9}$ in classical communication, the maximum tolerable phase noise is $\sigma_I+\sigma_B=0.0278$. This is determined from Eq.~(9) by requiring the denominator to be a real number.

\subsection{Secure key rate in QKD}

The asymptotic secure key rate of QKD, in the case of reverse reconciliation, is given by Refs.~\cite{Lodewyck07, Fossier09} 
\bes\label{eq11} R=fI_{AB}-\chi_{BE}, \ees
where $I_{AB}$ is the Shannon mutual information between Alice and Bob; $f$ is the efficiency of the reconciliation algorithm; $\chi_{BE}$ is the Holevo bound between Eve and Bob.

As we have discussed above, we assume detector noise $\upsilon_{el}$ and phase noise $\sigma_B$ are trusted, while phase noise $\sigma_I$, excess noise due to leakage $\varepsilon_{le}$ and channel noise $\varepsilon_0$ are untrusted. Under this noise model, in the case of conjugate homodyne detection, the detector-added noise referred to Bob's input is given by
\bes\label{eq12} \chi_{het}=(2+2\upsilon_{el})/\eta-1+T\varepsilon_B \ees
where $\varepsilon_B$ quantifies the excess noise due to the trusted phase noise referred to the channel input:
\bes\label{eq13} \varepsilon_B=(\dfrac{\alpha^2}{N_0}+V_A)\sigma_B. \ees

The total channel-added noise (including all the untrusted noise) referred to the channel input is given by
\bes\label{eq14} \chi_{line}=\frac{1}{T}-1+\varepsilon_{le}+\varepsilon_0+\varepsilon_I+\dfrac{4\alpha^2}{N_0}C_{BER} \ees
where $\varepsilon_I$ quantifies the excess noise due to untrusted phase noise and is given by $\varepsilon_I=(\dfrac{\alpha^2}{N_0}+V_A)\sigma_I$; The term $\dfrac{4\alpha^2}{N_0}C_{BER}$ quantifies the excess noise contributed by the BER in classical communication.  

The overall noise referred to the channel input is given by
\bes\label{eq15} \chi_{tot}=\chi_{line}+\dfrac{\chi_{het}}{T}. \ees

Since both quadratures are used for secure key generation, the mutual information between Alice and Bob is given by
\bes\label{eq16} I_{AB}=log_2\dfrac{V_A+1+\chi_{tot}}{1+\chi_{tot}}. \ees

The Holevo bound of the information between Eve and Bob is given by Ref.~\cite{Lodewyck07} 
\bes\label{eq17} \chi_{BE}=\sum_{i=1}^2 G\left( \dfrac{\lambda_i-1}{2} \right)  - \sum_{i=3}^5 G\left( \dfrac{\lambda_i-1}{2}\right),  \ees
where $G(x)=(x+1){\rm{log}}_2(x+1)-x{\rm{log}}_2x$

\bes\label{eq18} \lambda_{1,2}^2=\frac{1}{2} \left[ A\pm \sqrt{A^2-4B} \right] \ees
where
\bes\label{eq19} A=V^2 (1-2T)+2T+T^2 (V+\chi_{line})^2 \ees
\bes\label{eq20}B=T^2(V\chi_{line}+1)^2 \ees
 
\bes\label{eq21} \lambda_{3,4}^2=\frac{1}{2} \left[ C\pm \sqrt{C^2-4D} \right] \ees
where
\begin{equation}
\begin{split}
C=\dfrac{1}{(T(V+\chi_{tot}))^2} [ A\chi_{het}^2+B+1+2\chi_{het} \\
( V\sqrt{B}+T(V+\chi_{line})) +2T(V^2-1)]
\end{split}
\end{equation}

\bes\label{eq22}D=\left( \dfrac{V+\sqrt{B}\chi_{het}}{T(V+\chi_{tot})} \right) ^2 \ees 
\bes\label{eq23} \lambda_5=1. \ees

\subsection{Simulation results}

\begin{figure}[t]
	\includegraphics[width=.5\textwidth]{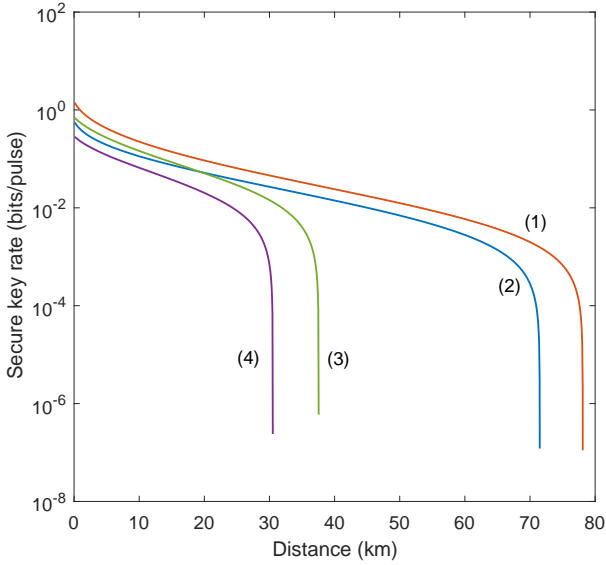}
	\captionsetup{justification=raggedright,
					singlelinecheck=false }
	\caption{Simulation results of secure key rate under the constraint of $10^{-9}$ BER in the classical communication. Simulation parameters: $\gamma=0.2$ dB/km, $\varepsilon_0=0.01$, $\upsilon_{el}=0.1$, $\eta=0.5$, $f=0.95$, $\Delta t=50ns$, $\tau_c=1\mu s$, $\xi_A=\xi_P=30dB$, and $n_{ref}=1000$. The modulation variance $V_A$ is numerically optimized at each fiber length. The four curves presented correspond to the following phase noise combinations: (1) $\sigma_I=10^{-5}, \sigma_B=10^{-3}$; (2) $\sigma_I=10^{-5}, \sigma_B=10^{-2}$; (3) $\sigma_I=10^{-4}, \sigma_B=10^{-3}$; and (4) $\sigma_I=10^{-4}, \sigma_B=10^{-2}$. As a comparison, using the above system parameters, no secure key can be generated at any distance if phase noise $\sigma_B$ is untrusted.}
	\label{fig:3}
\end{figure}

We conduct numerical simulations of the secure key rate of QKD under the constrain of $10^{-9}$ BER in the classical communication. Other simulation parameters are: $\gamma=0.2$ dB/km, $\varepsilon_0=0.01$, $\upsilon_{el}=0.1$, $\eta=0.5$, $f=0.95$, $\Delta t=50ns$, $\tau_c=1\mu s$, $\xi_A=\xi_P=30dB$, and $n_{ref}=1000$. At each distance, the channel transmittance $T$ can be determined using (7). From (9) and (10), given $T$ and $C_{BER}=10^{-9}$, the displacement $\alpha$ is determined by $V_A$ and other system parameters. So the only free parameter needs to be optimized is the modulation variance $V_A$. We numerically optimize $V_A$ at each distance to achieve the maximum secure key rate. Secure key rates are calculated at 4 different phase noise combinations: 1) $\sigma_I=10^{-5}, \sigma_B=10^{-3}$; 2) $\sigma_I=10^{-5}, \sigma_B=10^{-2}$; 3) $\sigma_I=10^{-4}, \sigma_B=10^{-3}$; and 4) $\sigma_I=10^{-4}, \sigma_B=10^{-2}$. Fig.~\ref{fig:3} shows the simulation results. As a comparison, we also calculate the secure key rate under the assumption that phase noise $\sigma_B$ is untrusted. Using the same system parameters, no secure key can be generated at any distance.
 
\section{Phase noise measurement}
\label{sec:4}

We conduct an experiment to determine the phase noise in a coherent comuunication system using a true LO based on the phase receovery shceme proposed in \cite{Marie17}. A similar experiment has also been conducted recently by Wang, et al., \cite{Wang17}. 

The experimental setup is shown in Fig.~4. Two commercial frequency-stabilized cw lasers at telecom wavelength (Clarity-NLL-1542-HP from Wavelength Reference) are employed as the signal laser and the LO laser. Both lasers are operated at free-running mode with no optical or electrical connections between them. Two LiNbO$_3$ waveguide amplitude modulators (EOSpace) are used to generate $10$ns laser pulses at a repetition rate of $10$MHz. At Alice's side, a polarization-maintaining fiber interferometer with a time-delay unbalance of 46.9ns is employed to generate a phase-related pulse pair (signal and phase reference) from each incoming pulse. A specially designed bias-free amplitude and phase modulator (APM in Fig.~4) is placed inside the interferometer to control the amplitude and phase of the signal pulse. Details of its design is presented in Fig.~5. Note the signal pulse and the phase reference pulse are coupled into orthogonal polarization modes by using a polarization beam combiner (PBC$_1$ in Fig.~4). Such a design can improve the isolation between the two pulses. Both the signal pulse and the phase reference pulse propagate through a spool of 25km single mode fiber. At Bob's end, a commercial $90^o$ optical hybrid (Optoplex) and two $350$ MHz balanced amplified photodetectors (Thorlabs) are employed to measure both X-quadrature and P-quadrature of the two pulses from Alice. The two LOs used in the coherent detection are split from a common pulse generated by the LO laser. A tunable optical delay line (TDL in Fig.~4) is placed inside Bob's interferometer to match its time delay difference with that of Alice's interferometer. By adjusting two polarization controllers (PC$_2$ and PC$_3$ in Fig.~4), the intensity of each LO pulse can be adjusted individually. Similar to the signal pulse and the reference pulse from Alice, the two LO pulses are also coupled into orthogonal polarization modes by a polarization beam combiner (PBC$_2$ in Fig.~4). Another polarization controller (PC$_1$ in Fig.~4) is used to match the polarization of Alice's pulse with that of the corresponding LO. Finally, the outputs of the two balanced photodetectors are sampled by a 12-bit data acquisition board (Texas Instruments).

\begin{figure}[t]
	\includegraphics[width=.5\textwidth]{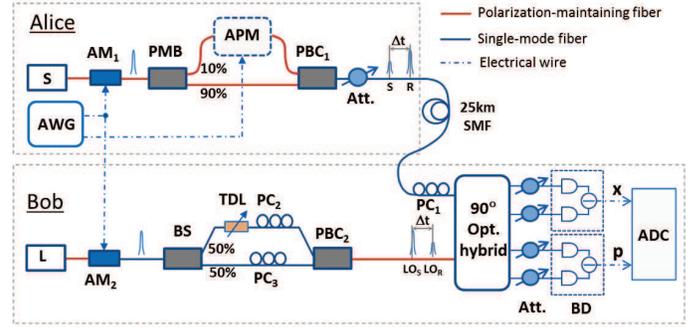}
	\captionsetup{justification=raggedright,
					singlelinecheck=false }
	\caption{Experimental setup. S-signal laser; L-LO laser; AM$_{1,2}$-amplitude modulator; APM-amplitude and phase modulator (see details in Fig.~5); PMB-90:10 polarization maintaining fiber beam splitter; BS-50:50 single mode fiber beam splitter; TDL-tunable optical delay line; PBC$_{1,2}$-polarization beam combiner; PC$_{1-3}$- polarization controller; Att.-tunable optical attenuator; AWG-arbitrary waveform generator; BD-balanced photodetector; ADC-analog-to-digital converter. } 
	\label{fig:4}
\end{figure}

In high-speed QKD, LiNbO$_3$ waveguide modulators are commonly employed to implement amplitude and/or phase modulation. The bias voltage control is vital for an amplitude modulator since its bias point commonly drifts with time. Here, we achieve bias-free amplitude and phase modulation by placing a phase modulator asymmetrically inside a loop interferometer, as shown in Fig.~5. See a simialr scheme in \cite{LDB96}. The basic idea is to introduce different phase shifts on lights traveling through the loop clock-wisely (CW pulse) or counter-clock-wisely (CCW pulse). Note, depending on the input optical signal (cw or pulsed) and the waveform of the electrical control signal on the phase modulator, this device can act as either an optical pulse generator, or an amplitude and phase modulator. When the input is cw light, each control pulse on the phase modulator will generate a pair of output light pulses with a time delay given by $\Delta t=n(L_2-L_1)/c$, where $(L_2-L_1)/2$ is the offset of the phase modulator from the middle point of the loop interferometer (which is about 1.8m is our experiment), $n$ is the refractive index of optical fiber, and $c$ is the speed of light in vacuum (see details in Fig.~5(a)). The temporal width of the output optical pulse is determined by the width of the control signal. When the input is pulsed light, bias-free amplitude and phase modulation can be achieved by controlling the waveform of the control signal to the phase modulator, as shown in Fig.~5(b). This design can be useful in other applications beyond QKD. In this experiment, we simply use it to adjust the photon number of signal.

\begin{figure}[t]
	\includegraphics[width=.45\textwidth]{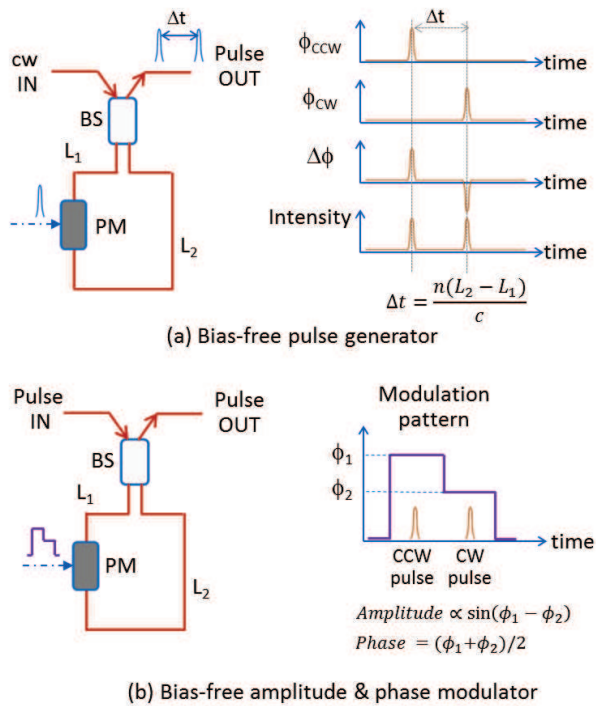}
	\captionsetup{justification=raggedright,
					singlelinecheck=false }
	\caption{Bias-free amplitude and phase modulator. BS-50:50 polarization maintaining fiber beam splitter; PM-phase modulator; CW-clockwise; CCW-counterclockwise. (a)When the input is cw light, the setup can act as a pulse generator; (b) When the input is pulsed light, the setup can act as an amplitude and phase modulator.} 
	\label{fig:5}
\end{figure}

In CV-QKD using a true LO \cite{Qi15}, the measurement results of the phase reference pulse $(X_{ref},P_{ref})$ are used to determine the phase difference between the signal laser and the LO laser using the relation
\bes\label{eq24} \phi=-tan^{-1}\dfrac{P_{ref}}{X_{ref}}.\ees

Once $\phi$ has been determined, Bob can correct his measurement results of the signal pulse by using (3). Here we want to determine the phase noise of the above process. More specifically, we want to quantify the difference between the $\phi$ estimated by Bob and the \emph{true} value of the phase difference $\phi_{tru}$ between the two lasers when the signal is measured. To acquire a precise estimation of $\phi_{tru}$, we replace Alice's signals for the SQCC protocol QKD by strong (unmodulated) calibration pulses. In fact, to minimize the measurement noise associated with the calibration pulse, their intensity is even stronger than that of the phase reference pulse. In this experiment, we define the weak pulse going through the path with an amplitude and phase modulator (see Fig.~4) as the phase reference pulse, and the strong pulse going through the other path as the calibration pulse. We remark that no information is encoded in this experiment.

From the measurement results of the calibration pulse $(X_{c},P_{c})$, we calculate 
\bes\label{eq25} \phi_{tru}=-tan^{-1}\dfrac{P_c}{X_c}.\ees

The phase error is defined as $\phi-\phi_{tru}$. Experimentally, the variance of $\phi-\phi_{tru}$ has been determined to be 2.4$\pm$0.4$\times 10^{-3} rad^2$ (when the average photon number of the phase reference pulse is $10^3$), and 0.79$\pm$0.25$\times 10^{-3} rad^2$ (when the average photon number of the phase reference pulse is $10^4$). From Fig.~3, this phase noise is low enough to implement the SQCC protocol over practical distances.

\section{Discussion}
\label{sec:5}

CV-QKD based on optical coherent detection is appealing in practice since it can be implemented with standard telecommunication technology \cite{Diamanti15}. The research in CV-QKD is also aligned with the resurgence of classical optical coherent communication, which is the most promising solution to the dramatic growth of global communication traffic \cite{Agrell16}. Studies in this paper show it is feasible to use the same coherent communication system to conduct both QKD and classical communication simultaneously, as long as the distance is within the reach of QKD.

To improve phase noise tolerance of the SQCC protocol and make it compatible with the CV-QKD scheme using a true LO, in this paper we adopt a refined noise model where phase noise due to finite photon number of the phase reference pulse and the detector imperfection is assumed to be trusted. Systematic noise analysis has been conducted. Simulation results show that the tolerable phase noise in this refined noise model is significantly higher than that in previous studies~\cite{Qi16}. Experimentally, using a design proposed in Ref.~\cite{Marie17}, we demonstrated the required phase stability can be achieved in practice.

While the results presented in this paper are encouraging, further researches are needed to bring this technology into real life. As we have shown in this paper, a trusted noise model could significantly improve the QKD performance. However, it could also introduce potential security loopholes if Eve has a way to manipulate the phase noise, or the QKD users overestimate the amount of trusted phase noise. It is thus important to implement local calibration systems at both Alice and Bob to monitor the relevant noises in real time.

We thanks the anonymous reviewers for their important comments on the trusted noise model and the noise associated with the leakage of the phase reference pulses.

This work was performed at Oak Ridge National Laboratory (ORNL), operated by UT-Battelle for the U.S. Department of Energy under Contract No. DE-AC05-00OR22725. The authors acknowledge support from ORNL Laboratory Directed Research and Development Program. CCW Lim acknowledges support from MOE AcRF grant R-263-000-C78-133/731.

\appendix

\section{Leakage from phase reference pulse}

In this Appendix, we study the leakage from the phase reference pulse to the signal based on the specific design presented in Section IV.

As shown in Fig.~4, Alice generates laser pulses from a cw laser source by using an optical amplitude modulator. Each laser pulse is further split into two (phase reference pulse and signal pulse) by using a path-unbalanced interferometer. The signal and the reference are coupled into orthogonal polarization modes to improve the isolation between them.

In practice, only finite extinction ratio can be achieved in both amplitude modulation and polarization multiplexing. So there will be unavoidable leakage from the phase reference pulse to the signal pulse, as highlighted in Fig.~6. 

\begin{figure}[t]
	\includegraphics[width=.45\textwidth]{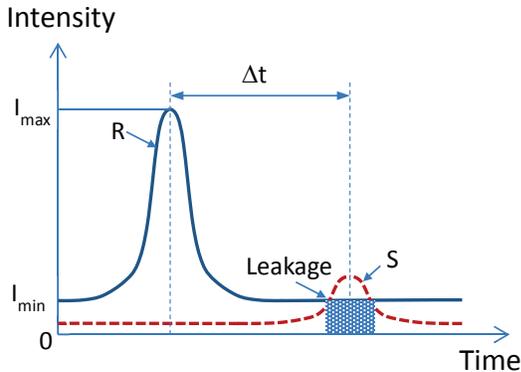}
	\captionsetup{justification=raggedright,
					singlelinecheck=false }
	\caption{Leakage from the phase reference pulse to the signal pulse due to the finite extinction ratio. R-phase reference pulse; S-signal pulse; $\Delta t$-time delay introduced by the path-unbalanced interferometer.} 
	\label{fig:6}
\end{figure}

Given the average photon number of the phase reference pulse at Bob's end is $n_{ref}$, the average photon number of the leakage at Alice's end is determined by
\bes\label{eq26} n_{le}=\dfrac{n_{ref}}{T}\times 10^{-\frac{\xi_A}{10}}\times 10^{-\frac{\xi_P}{10}},\ees
where $\xi_A$ and $\xi_P$ are the extinction ratios (in dB) of the amplitude modulator and the polarization multiplexing scheme, correspondingly.

If the leakage photon has a fixed phase relation with the phase reference pulse, then it only introduces a constant displacement in phase space, which can be determined from Bob's measurement results and removed in the post-processing stage \cite{Qi07}. When taking into account the phase noise of the QKD system and the finite coherent time of the signal laser, the excess noise contributed by the leakage can be described by
\bes\label{eq27} \varepsilon_{le}=\dfrac{n_{le}}{2N_0} \sigma_{le},\ees
where $\sigma_{le}$ quantifies the phase noise of the leakage.

In our setup, $\sigma_{le}$ is mainly determined by the coherent time $\tau_c$ of the laser. Note the phase reference pulse and the leakage are emitted at different times (with a time delay of $\Delta t$) by Alice's laser. The spontaneous emitted photons generated
within the above time interval contribute a fundamental phase noise with a variance of $2\Delta t/\tau_c$. As shown in \cite{Qi15}, the coherent time of our laser is about 1 $\mu s$. If $\Delta t$ is about 50ns (see Fig.~6), then the phase noise of the leakage contributed by the laser is about 0.1 $rad^2$, which is much larger than other phase noises in the QKD system.

Using $\sigma_{le}=2\Delta t/\tau_c$ and (A1), Eq.(A2) can be revised as
\bes\label{eq28} \varepsilon_{le}=\dfrac{n_{ref}\Delta t}{T N_0 \tau_c} \times 10^{-\frac{\xi_A}{10}}\times 10^{-\frac{\xi_P}{10}}.\ees

\end{document}